\documentclass[a4paper,fleqn,usenatbib]{mnras}

\usepackage{newtxtext,newtxmath}

\usepackage[T1]{fontenc}
\usepackage{ae,aecompl,ulem}
\usepackage{amstext}
\usepackage{relsize, float, multirow}
\usepackage{array, epstopdf, hyperref, }
\usepackage{caption, subcaption, mathtools}
\usepackage{newtxmath}

\DeclareMathOperator{\arcsinh}{arcsinh}

\usepackage{graphicx}	
\usepackage{amsmath}	

\def\aap{A\&A}
\def\apj{ApJ}

\def\apjl{ApJL}
\def\mnras{MNRAS}

\def\nat{Nature}

\def\physrep{Physics Reports}         
\def\pasj{PASJ}              
\def\apss{ApSS}
\def\jcap{JCAP}
\def\sovast{Soviet Astronomy}

\def \tcr{\textcolor{red}}

\title[Non-thermal SZ from radio galaxy lobes]{Non-thermal Sunyaev-Zeldovich signal from radio galaxy lobes} 

\author[Acharya, Majumdar \& Nath]{
Sandeep Kumar Acharya,$^{1}$\thanks{sandeepkumar@theory.tifr.res.in}
Subhabrata Majumdar,$^{1}$\thanks{subha@tifr.res.in}
Biman B. Nath$^{2}$\thanks{biman@rri.res.in}
\\
$^{1}$Department of Theoretical Physics, Tata Institute of 
Fundamental Research, Mumbai 400005, India\\
$^{2}$Raman Research Institute, Sadashiva Nagar, Bangalore 560080, India
}

\date{Accepted XXX. Received YYY; in original form ZZZ}

\pubyear{2020}

\begin{document}
\label{firstpage}
\pagerange{\pageref{firstpage}--\pageref{lastpage}}
\maketitle

\begin{abstract}
Energetic electrons in the lobes of radio galaxies make them potential sources for not only radio and X-rays but also Sunyaev-Zeldovich (SZ) distortions in the cosmic microwave background (CMB) radiation. Previous works have discussed the energetics of radio galaxy \color{black}lobes \color{black}, but assuming thermal SZ effect \color{black}, coming from the non-thermal electron population. We use an improved evolutionary model for radio galaxy \color{black}lobes \color{black} to estimate the observed parameters such as the radio luminosity and intensity of SZ-distortions at the redshifts of observation. We, further, quantify the effects of various relevant physical parameters of the radio galaxies, such as the jet power, the time scale over which the jet is active, the evolutionary time scale for the \color{black}lobe \color{black}, etc on the observed parameters. For current SZ observations towards galaxy clusters, we find that the non-thermal SZ distortions from radio \color{black}lobes \color{black} embedded in galaxy clusters can be non-negligible  compared to the amount of thermal SZ distortion from the intra-cluster medium and, hence, can not be neglected. We show that small and young (and preferably residing in a cluster environment) radio galaxies offer better prospects for the detection of the non-thermal SZ signal from these sources. We further discuss the limits on different physical parameters for some sources for which SZ effect has been either detected or upper limits are available. The evolutionary models enable us to obtain limits, previously unavailable, on the low energy cut-off of electron spectrum  ($p_{min} \sim 1\hbox{--}2$) in order to explain the recent non-thermal SZ detection. 
Finally, we discuss how future CMB experiments, which would cover higher frequency bands ($>$400 GHz), may provide clear signatures for non-thermal SZ effect.
\end{abstract}




\section{\label{sec:intro}Introduction}
Studying the distortion of the cosmic microwave background radiation (CMB) through the  Sunyaev-Zel'dovich effect (SZ \color{black}effect\color{black}) \citep{ZS1969} has become a mainstay of 
modern cosmology. The inverse Compton scattering of the CMB photons by energetic electrons has opened up a new window of probing the warm and hot gaseous regions of the universe \citep{B1999,AMS2008,M2019}. Being complementary to the traditional X-ray observation of these ionized regions, SZ \color{black}effect \color{black} not only enables a robust determination of their physical properties,
but also makes it possible to study them at high redshift, and consequently, study their evolution. The intensity of distortion of CMB does not dilute with increasing redshift resulting from the fact that the energy density of scattered CMB photons and the intensity of CMB photons have same functional dependence on redshift.


Besides energetic thermal electrons in hot gas, non-thermal relativistic gas can also  produce SZ \color{black}signal\color{black}, whose distinct spectral signature makes it an interesting probe 
of reservoirs of such gas in the universe \citep{EK2000,M2001}. One source of  energetic particles can be radio galaxy \color{black}lobes \color{black} where relativistic particles are supplied by the radio jet \citep{S1974,BC1989,N1995},
which was first predicted by \cite{FR1969} soon after the discovery of CMB. The pressure from the energetic particles can push out the surrounding gas with the size of the \color{black}radio lobe \color{black} growing to megaparsec length scales \citep{B1982,KDA1997}. 
Recent detection of X-ray emission from radio lobes has been explained through inverse Compton scattering of CMB photons by non-thermal relativistic electrons that are responsible for the radio emission \citep{CHHBBW2005,EFB2008,FCCBB2009,JABD2007,ITGHNHSM2009}. These studies have utilized the fact that relativistic non-thermal electrons are the source for both the X-ray and the radio emission, and can therefore deduce the magnetic fields in these objects with greater confidence than were previously possible \citep{ITGHNHSM2009}. At the same time, the inverse Compton X-ray points towards a concomitant distortion in the CMB.
\par
 The non-thermal SZ effect from radio galaxies was studied by \cite{C2008} by inferring the shape of electron spectrum from the observed X-ray signal. This idea was further extended by \cite{CM2011} by using X-ray and radio signal from radio galaxies. The authors concluded that non-thermal SZ signal gives a reliable understanding of the low energy cutoff of the electron spectrum to which X-ray and radio signals are insensitive.  Following this idea, there have been searches for non-thermal SZ \color{black} signal \color{black} towards radio galaxies, which have fetched upper limits \citep{CMBM2013} 
until a recent detection \citep{MDCMSNW2017}. The shape of this distortion is a function of non-thermal spectrum of these relativistic particles \citep{B1999,EK2000} and is distinct from the shape of the thermal $y$-distortion \citep{ZS1969}. 
 In previous works,\cite{YSS1999} \& \cite{M2001} had estimated the global $y-$distortion caused by radio galaxies to be $\approx \ge 10^{-5}$, using a simple model of radio \color{black}galaxy \color{black} evolution and cosmological population of radio galaxies constructed from the Press-Schechter mass function. 
  
In this paper, we first present an improved model for the evolution of radio \color{black}galaxies \color{black} in order to predict the non-thermal SZ signal from radio galaxy \color{black}lobes \color{black} which can be targets for upcoming SZ observations. We use the galaxy evolution model of \cite{KDA1997} (KDA1997 hereafter) with suitable modification for jet stopping as in \cite{N2010} (N2010 hereafter). We explicitly keep track of the  evolution of the relativistic particles. We then calculate the non-thermal SZ spectrum from these population of relativistic particles. Using the procedure detailed in this work, future observations of non-thermal SZ effect can be used to put constraints on the underlying physical model of radio galaxies, for example, that on the jet luminosity and lifetime, as also on the nature of relativistic population of electrons driven out by the radio jet, for example, the lowest energy threshold of the electron spectrum. The rest of the paper is arranged as follows. In Sec \ref{sec:model} we lay down our model for the radio galaxy \color{black}lobes \color{black}, followed by physics related to SZ distortions from non-thermal relativistic population of electrons in Sec \ref{sec:nonthSZ}. These are combined to estimate the non-thermal SZ from radio \color{black}galaxy lobes \color{black} in Sec \ref{sec:SZcocoon}. Next, Sec \ref{sec:degeneracy} \& \ref{sec:detection} discuss the degeneracies in model parameters and the prospects of future detection of non-thermal SZ effect. We discuss our results in \ref{sec:discuss} and, finally, conclude in \ref{sec:conclude}. For calculating the angular diameter distance in converting angular to physical size we use $\it{H_0}$ = 67 km s$^{-1}$ Mpc$^{-1}$, $\it{\Omega_M}$ = 0.32, $\it{\Omega_\Lambda}=0.68$ \citep{Pl2018}.
    
    \begin{figure}
\includegraphics[width=\columnwidth]{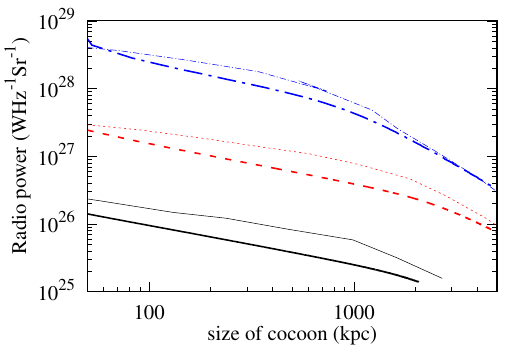}
\caption{Comparison of radio power, at 150 MHz, obtained in this work (in thick lines) with the corresponding values in KDA1997 (thin lines) for the three cases as given in KDA1997 (their Fig. 1). The parameters for the blue (dot-dashed), red (dashed) and black (solid) lines correspond to [$\it{Q_{\rm{J}}}$ (erg s$^{-1}$),\,$\it{z}$]\,=\,[($10^{47},\,2$ ), ($10^{46},\,0.5$ ) \& ($10^{45},\,0.2$)]  respectively where $\it{Q_{\rm J}}$ is the jet luminosity and $\it{z}$ is the observed redshift of radio galaxy.The spectral index of electron energy is taken to be $\rm{\alpha_p}=2.15$.} 
\label{KDA} 
\end{figure}
    
\section{Model for radio galaxy lobes}
\label{sec:model}
 We start by assuming that the jet of the radio galaxy, which has a luminosity $\it{Q_{\rm J}}$, injects relativistic particles throughout the jet lifetime $\it{t_{\rm J}}$ ($\approx 10^{7\hbox{--}8}$ yr), after which it ceases to be active. 
 The injected energy causes a \color{black}lobe \color{black} around the jet to expand against the surrounding medium. 
  We describe the density profile of the surrounding gas with a power law, ${\rho(r)=\rho_0(\frac{r}{a_0})^{-\beta_g}}$ with \color{black}${\beta_g}$=2 \color{black}\citep{WK2008}. This density profile can be written as, ${\rho(r)=\Lambda r^{-2}}$, with ${\Lambda}=10^{19}$g cm$^{-1}$ \citep{FMO2004,JPHC2007}. The non-thermal electron energy distribution \color{black} in the radio lobe \color{black} is assumed to be a power law with, $n(\gamma_{\rm{i}},t_{\rm{i}})=n_0\gamma_{\rm{i}}^{-\alpha_p}d\gamma_{\rm{i}}$, where $\gamma_{\rm{i}}$ is the Lorentz factor of electrons at time of injection $t_{\rm{i}}$, and ${\alpha_p}$ is the spectral index. We assume the minimum ${\gamma_{\rm{min}}=1}$ 
  and the maximum, ${\gamma_{\rm{max}}=10^6}$. The evolution of the radio \color{black}galaxy  lobes \color{black} can be described by \citep{RB1997},
  \begin{equation}
  {Q_{\rm J}(t)=\frac{1}{\Gamma_c-1}(V_c \overset{.}{p_c}+\Gamma_c p_c\overset{.}{V_c}) \,, \quad
 \frac{dL_{\rm J}}{dt}=\left(\frac{p_c}{\rho}\right)^{1/2}},
 \label{cocoonsize}
 \end{equation}
 where ${Q_{\rm J}(t)}$ is the jet luminosity which is non-zero when jet is on (${t<t_{\rm J}}$, where ${t_{\rm J}}$ is the jet lifetime) and zero for ${t>t_{\rm J}}$, ${\Gamma_c=\frac{4}{3}}$, ${V_c}$ is the volume of the \color{black}radio lobe \color{black}, ${p_c}$ is the pressure in the \color{black}radio lobe \color{black}, ${L_{\rm J}}$ is the fiducial size of the \color{black}lobe  \color{black} and ${\rho}$ is the density of surrounding gas. We assume the axial ratio of the cylinder shaped \color{black}radio lobe \color{black} to be $\it{R=2}$, the average observed ratio \citep{LW1984}. The volume of the \color{black}lobe \color{black} is then given by,
 \begin{equation}
 {V_c=\frac{\rm \pi}{4R^2}L_{\rm J}^3} \,.
 \end{equation}
 The magnetic and particle energy densities in the \color{black}lobe \color{black} are given by,
 \begin{equation}
 {U_B(t)=\frac{Ap_c(t)}{(\Gamma_c-1)(1+A)} \,, \quad U_e(t)=\frac{p_c(t)}{(\Gamma_c-1)(1+A)}} \,,
 \end{equation} 
 where ${A=(1+\alpha_p)/4}$ \citep{K1997}.
 For a comparison with result of KDA1997, we first consider the case when jet is on all the time. We then proceed to compute the radio flux density at 150 MHz as a function of time or size of the \color{black}radio lobe \color{black}. Assuming that for synchrotron radiation, an electron emits only at the frequency ${\nu=\gamma^2 \nu_L}$, where $\nu_L$ is the Larmor frequency, we calculate the number density of electrons with Lorentz factor ${\gamma_{150 \rm{MHz}}}$ which will emit at 150 MHz as a function of time $\rm{t}$. We then find the value of $\gamma_{\rm{i}}$ which were injected at time $\rm{t_i<t}$ and which would have cooled to ${\gamma_{150 \rm{MHz}}}$ at time t. The equation for evolution of electron  Lorentz factor ${\gamma}$ is given by,
 \begin{equation}
{\frac{d\gamma}{dt}=-\frac{1}{3}\frac{1}{V_c}\frac{dV_c}{dt}-\frac{4}{3}\frac{\it{\sigma_{\rm{T}}}}{\it{m_{\rm{e}} \rm{c}}}\gamma^2 (U_B+U_C)}\,,
\label{cooling}
\end{equation}
where $\sigma_{\rm{T}}$ is Thomson cross-section, $m_{\rm{e}}$ is the mass of electron, ${c}$ is the speed of light, $U_B$, $U_C$ are the magnetic energy density and CMB energy density, respectively. \color{black} This equation assumes no re-acceleration of electrons (since the RHS is negative). \color{black} The normalization of particle spectrum ${n_0}$ at time $t_{\rm{i}}$ is given by \citep{KDA1997,N2010},
\begin{equation}
n_0(t_{\rm{i}})=\frac{U_e(t_{\rm{i}})}{m_{\rm{e}}{c}^2}\int_{\gamma_{\rm{min}}}^{\gamma_{\rm{max}}}(\gamma_{\rm{i}}-1)\gamma_{\rm{i}}^{-\alpha_p}d\gamma_{\rm{i}} \,.
\end{equation}  
The number density of electrons with $\rm{\gamma=\gamma_{150 \rm{MHz}}}$, at time $\rm{t}$ due to the expansion of \color{black}radio lobe \color{black}, is given by,
\begin{equation}
n(\gamma)=n_0\frac{\gamma_{\rm{i}}^{2-\alpha_p}}{\gamma^2}\left(\frac{p_c(t_{\rm{i}})}{p_c(t)}\right)^{-1},
\label{numberdensity}
\end{equation} 
where $\frac{1}{V_c}\frac{dV_c}{dt}$=$\frac{3}{L_{\rm J}}\frac{dL_{\rm J}}{dt}$ and the \color{black}radio lobes \color{black} are assumed to evolve self-similarly (as warranted by the assumption 
of a constant axial ratio). Any volume segment of the \color{black}radio lobe \color{black} at time $t$ can be related to pressure at time $t_i$ as,
\begin{equation}
\delta V(t)=\frac{(\Gamma_c-1)Q_{\rm J}}{p_c(t_{\rm{i}})}(4R^2)^{(1-\Gamma_c)/\Gamma_c}\left(\frac{p_c(t)}{p_c(t_{\rm{i}})}\right)^{1/\Gamma_c}\delta t_{\rm{i}},
\end{equation}
where $\delta t_{\rm{i}}$ is the time interval over which electrons were injected. Then, the power emitted at $t$ at 150 MHz is given by,
\begin{equation}
P_{\nu}=\int_0^t\frac{1}{6 \rm{\pi}}\sigma_{\rm{T}}\rm{c}U_B\frac{\gamma_{150 \rm{MHz}}^3}{\nu}n(\gamma_{150 MHz}) \delta V,
\end{equation}
where the integral is done over electron injection time $t_{\rm{i}}$.
For jet shutdown at time $t_{\rm J}$, the upper limit of integral should be min($t,t_{\rm J}$) \citep{N2010}.

\begin{table}
    \begin{tabular}{l|c|c|c|r} 
      $x$ & $p_{min}=1$ & 3 & 5 &10   \\
      \hline
      0.47 (27 GHz) & -0.16 & -0.028 & -0.011 & -0.003 \\
      0.7 (39 GHz) & -0.33 & -0.055 & -0.021 & -0.005 \\
      1.7 (93 GHz) & -1.08 & -0.18 & -0.068 & -0.018 \\
      2.54 (145 GHz) & -1.3 & -0.22 & -0.086 & -0.023 \\
       4.1 (225 GHz) & -0.72 & -0.17 & -0.069 & -0.019 \\
       5.0 (280 GHz) & -0.2 & -0.1 & -0.05 & -0.013 \\
    \end{tabular}
   \caption{Value of $g_{NT}(x)$ as in Fig. \ref{gx} at the frequency band of Simons Observatory \citep{SO2019}.}
   \label{tab:table1}
\end{table}

In Fig. \ref{KDA}, we have compared our results with the result of KDA1997 with the jet on for all the time for three different cases. The differences between the solid (this work) and dashed (KDA1997) lines are due to the fact that we have used equation \ref{cocoonsize} instead of using a power-law solution ($L_{\rm J} \propto t^{3/5-\beta}$) at all times.
A higher jet luminosity increases the radio power due to a larger number of energetic particles. This, also, increases the size of \color{black} the lobe \color{black} due to an increase in pressure owing to these particles. The initial expansion of the \color{black}lobe \color{black} is dominated by the pressure of the energetic particles and the synchrotron and inverse Compton cooling can be ignored. However, once the pressure drops due to expansion of the \color{black}lobe \color{black}, the inverse Compton cooling becomes important. At higher redshifts, cooling by CMB photons become increasingly efficient as CMB energy density is proportional to $(1+z)^4$. This can be seen in the leftward shift in the break of slope of the curve at  $\sim$ 1000 kpc since there are less number of energetic electrons at  $\gamma_{\rm{150 MHz}}$ due to efficient cooling. 

\section{Sunyaev-Zeldovich effect from non-thermal relativistic population of electrons}
\label{sec:nonthSZ}
\begin{figure}
\centering
\includegraphics[width=\columnwidth]{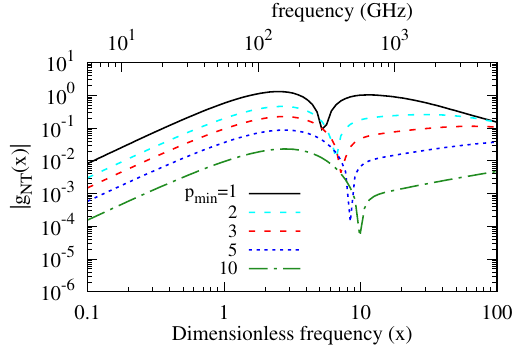}
\caption{ \color{black}\color{black} Absolute value of spectral function $g_{\rm{NT}}(x)$ for power law distribution of electrons number density ($f_{\rm{e}}(p)\propto p^{-\alpha_p}$) with $\alpha_p=3$. The lowest momentum cutoff is denoted by $p_{\rm{min}}$ and the highest momentum cutoff is chosen to be $p_{\rm{max}} \sim10^6$. In magenta line, we show the $g_{\rm{NT}}(x)$ for $\alpha_p=2.3$ and $p_{\rm{min}}=1$. The cuspy feature is due to changing sign of $g_{\rm{NT}}(x)$ with $g_{\rm{NT}}(x)$ being negative in value to the left of the cusp and positive to the right of the cusp. \color{black}   Note, $p=\sqrt{(\gamma^2-1)}$.}  
\label{gx}
\end{figure}

It is well known that energetic electrons can boost the CMB photons to higher energy through inverse Compton scattering creating a distortion in the CMB blackbody spectrum. If the energy distribution of the electrons is non-relativistic and thermal, then the distortion has \color{black} $y$-distortion shape \citep{ZS1969} (See Sec. \ref{sec:detection} for further discussions on corrections to $y$-distortion)  \color{black}. For relativistic electrons, with a Lorentz factor $\gamma$, a photon with energy $\epsilon$ gets boosted to $\gamma^2\epsilon$. In this case, the spectral distortion shape will be a function of the electron energy distribution. 
The intensity of the CMB spectrum per frequency is given by,
\begin{equation}
I_{\nu}(x)=2\frac{(k_{\rm{B}}T_{\rm{CMB}})^3}{({hc})^2}\frac{x^3}{e^x-1}=I_0I(x),
\end{equation}
where $I(x)=\frac{x^3}{e^x-1}$, $I(x)$ is the dimensionless intensity, $x$ is the dimensional frequency which is given by $x=\frac{E_{\gamma}}{k_{\rm{B}}T_{\rm{CMB}}}$,where $E_{\gamma}$ is the energy of photon, $k_{\rm{B}}$ is the Boltzmann constant, $T_{\rm{CMB}}$ is the CMB temperature, and other symbols have usual meanings. The intensity of distorted CMB spectrum is independent of redshift for a given population of electrons. The CMB distortion  in the optically thin limit can be written as \citep{ZS1969,B1999},
\begin{equation}
\Delta I(x)=(j(x)-i(x))\tau,
\label{delI}
\end{equation}
where $j(x)$ is the spectral intensity of photons at frequency $x$ after being upscattered while $i(x)$ is the intensity of photons at frequency $x$ before upscattering, $\tau=\sigma_{\rm{T}}\int n_{\rm{e}}dl$ where $n_{\rm{e}}$ is the electron number density and $dl$ is the line of sight width of this electron population. $i(x)$ is non-zero \color{black} over the range of $x$ where the CMB photons are present.  \color{black} Eq. \ref{delI} can be recast to include a $y$-parameter as $\Delta I(x)=yg(x)$, where $y=\frac{\sigma_{\rm{T}}}{m_{\rm{e}}\rm{c}^2}\int n_{\rm{e}}k_{\rm{B}}\overset{\sim}{T_{\rm{e}}}dl$ with $k_{\rm{B}}\overset{\sim}{T_{\rm{e}}}=\frac{P_{\rm{e}}}{n_{\rm{e}}}$. \color{black} Here $\overset{\sim}{T_{\rm{e}}}$ refers to a fictitious temperature scale for a non-thermal distribution to make the equations look suggestively similar to thermal SZ (more details can be found in \cite{EK2000})\color{black}. In order to distinguish non-thermal spectral distortion shape from $y$-type distortion (the well known thermal $\rm{SZ}$ \color{black}effect)\color{black}, we will refer to non-thermal distortion amplitude as $y_{\rm{NT}}$, such that,
\begin{equation}
\Delta I_{\rm{NT}}(x)=y_{\rm{NT}}g_{\rm{NT}}(x),
\label{ntSZ}
\end{equation}
where $g_{\rm{NT}}(x)$ is the spectral distortion function resulting from scattering of the CMB in a non-thermal population of clusters. \color{black}We use the formalism of \cite{EK2000} to compute $g_{\rm{NT}}(x)$ and we refer the reader to the paper for details \color{black}.
The pressure for a distribution of relativistic electrons is given by,
\begin{equation}
P_{\rm{e}}=n_{\rm{e}}\int dp f_{\rm{e}}(p)\frac{1}{3}p{v}(p)m_{\rm{e}}c,
\end{equation}
where $f_{\rm{e}}(p)$ is the normalized electron spectrum i.e. $\int f_{\rm{e}}(p)dp=1$ with \color{black} electron dimensionless momentum \color{black} $p=\sqrt{(\gamma^2-1)}$, ${v}=\beta c$, \color{black} where $\gamma$ is the Lorentz factor and $\beta$ is the boost factor of energetic electrons. \color{black}
     The number of CMB photons which get upscattered from energy $x'$ to $x$ is given by,
\begin{equation}
N(x'->x)= P(t,p)\times 2\frac{(k_{\rm{B}}T_{\rm{CMB}})^2}{({hc})^2}\frac{x'^2dx'}{e^{x'}-1}, 
\label{number}
\end{equation}
where $P(t,p)$ is the kernel of the inverse Compton scattering which captures the kinematics of photon scattering with the electrons with electron energy $p$, $t=\frac{x}{x'}$. The number of CMB photons within energy $x'$ and $x'+dx'$ is $2\frac{(k_{\rm{B}}T_{\rm{CMB}})^2}{({hc})^2}\frac{x'^2dx'}{e^{x'}-1}$ and $\int dt P(t,p)=1$, which conserves the number of photons. The formula for $P(t,p)$ is given by \citep{EK2000},
\begin{multline}
P(t;p)=\frac{-3\left|(1-t)\right|}{32p^6t}[1+(10+8p^2+4p^4)t+t^2]\\+\frac{3(1+t)}{8p^5}\left[\frac{3+3p^2+p^4}{\sqrt{1+p^2}}-\frac{3+2p^2}{2p}(2\arcsinh{p}-\left| \ln{t}\right|)\right],
\end{multline}
The spectral intensity of upscattered photons per frequency, in frequency bin $x$ and $x+\Delta x$, is given by,
\begin{equation}
j(x)=\frac{\int \int f_{\rm{e}}(p)dp P(t,p)\frac{x'^2dx'}{e^{x'}-1}x}{\Delta x}.
\label{j(x)}
\end{equation}
\par
In Fig. \ref{gx}, we plot the absolute value of spectral function $g_{\rm{NT}}(x)$ for power law distribution ($f_{\rm{e}}(p)\propto p^{-\alpha_p}$) with power law index \color{black}$\alpha_p$=3.0 and 2.3\color{black}. The minimum electron energy in the power law distribution is denoted by $p_{\rm{min}}$. \color{black} Higher $p_{\rm{min}}$ reduces $g_{\rm{NT}}(x)$ in the CMB spectral band (x<10) as CMB photons are scattered to $x>10$ energies efficiently by energetic electrons. Decreasing $\alpha_p$ while keeping $p_{\rm{min}}$ fixed also reduces the magnitude of distortion in the CMB band as there are more number electrons in the higher energy tail. \color{black} The value of $g_{\rm{NT}}(x)$ at specific frequencies corresponding to frequency band of the upcoming Simons Observatory \citep{SO2019} is given in Table \ref{tab:table1}.

\section{Non-thermal SZ signal from radio galaxy lobes}
\label{sec:SZcocoon}

\begin{figure}
\centering
\includegraphics[width=\columnwidth]{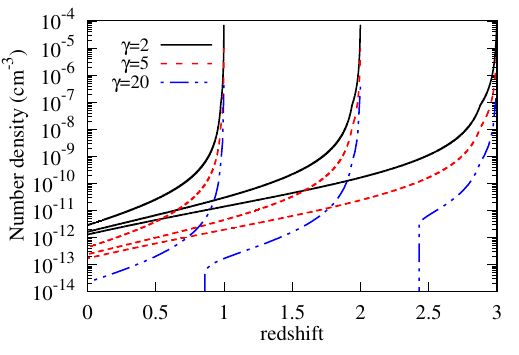}
\caption{Differential Number density of electrons within $\gamma$ and $\gamma+d\gamma$  inside a radio \color{black}lobe \color{black} for jets starting at different redshifts ($z= 3,\,2\, \& \,1$) for $Q_{\rm J}=10^{45}$ erg s$^{-1}$ and $t_{\rm J}=10^8$ yr. For each starting redshift, we show the results for three different energy distributions given for $\gamma$ = 2 (black solid lines),\, 5 (red dashed lines) and 20 (blue dot-dashed lines). See text for details regarding the abrupt fall of the $\gamma$=20 curves.} 
\label{ndens}
\end{figure} 

It is clear from the previous discussion that we need to calculate the number density of relativistic electrons as a function of the energy in order to calculate the non-thermal SZ spectrum for individual radio \color{black}lobe \color{black}. To proceed, we use the same strategy as used to calculate the number density of electrons which emit at 150 MHz. We divide the range in $\gamma$ from 1 to $10^6$ in 400 log spaced bins. Then, we use Eq. \ref{cooling} and \ref{numberdensity} for the whole range of electron energy (or $\gamma$) at each instant of time to calculate the number density $n(\gamma)d\gamma$ at that instant of time. 

In Fig. \ref{ndens}, we plot the number density of electrons with three different instantaneous energy with jets starting at different redshifts. The lowest energy electrons dominate the total number of relativistic particles. The cooling of electrons via inverse Compton is proportional to $\gamma^2U_C$. Therefore, at higher redshifts electrons cool much more efficiently via CMB photons. This can be clearly seen for $\gamma=20$ curve for a jet starting at redshift $2, 3$, as the electron number  precipitously drops after reaching $z=1, z=2.5$ respectively. The efficiency of cooling of electrons is a  strong function of redshift ($\propto (1+z)^4$). Therefore, the number density of electrons with the relatively higher $\gamma$ = 20  falls immediately after jet the is shut off at higher redshift (e.g. for jet starting at $z$=3)  while for jet starting at $z$=1, there are energetic electrons left even after the jet is shut off.

\begin{figure}
\centering
\includegraphics[width=\columnwidth]{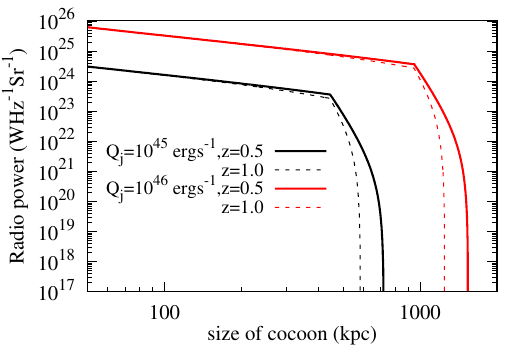}
\caption{
Radio power at \color{black} 150 MHz \color{black} as a function of the size of the \color{black}radio lobe \color{black} for a combination of jet luminosities $Q_{\rm J}$ and starting redshifts. The upper two red lines are
for higher $Q_{\rm J} = 10^{46}$ erg s$^{-1}$ and lower two black lines are for a lower $Q_{\rm J} = 10^{45}$ erg s$^{-1}$, with solid lines corresponding to $z_{st}\,=\,0.5$ and dashed lines for $z_{st}\,=\, 1$. The jet lifetime is $t_{\rm J}= 10^8 $ yr. \color{black} The rapid decline in radio power is due to jet stopping and no re-acceleration of electrons. \color{black}}  
\label{radiopowervsz}
\end{figure}

\begin{figure}
\centering
\includegraphics[width=\columnwidth]{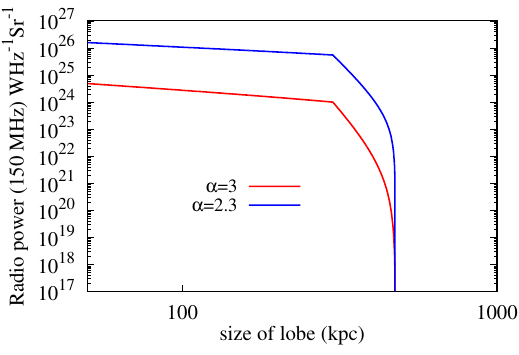}
\caption{
\color{black}Radio power at 150 MHz as a function of the size of the radio lobe for $\alpha_p=3$ and 2.3 respectively. Jet luminosity $Q_{\rm J}=10^{45}$erg s$^{-1}$, starting redshift $z_{st}=0.5$ and $t_{\rm J}=10^8$ yr.}  
\label{radiopowervsalpha}
\end{figure}

\begin{figure}
\centering
\includegraphics[width=\columnwidth]{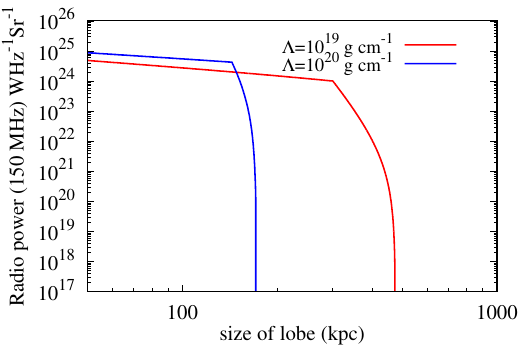}
\caption{
\color{black}Radio power at 150 MHz as a function of the size of the radio lobe for different density of surrounding gas with $\Lambda=10^{19}, 10^{20}$ g cm$^{-1}$ respectively. Jet luminosity $Q_{\rm J}=10^{45}$erg s$^{-1}$, starting redshift $z_{st}=0.5$ and $t_{\rm J}=10^8$ yr.}  
\label{radiopowervsdensity}
\end{figure}

\begin{figure}
\centering
\includegraphics[width=\columnwidth]{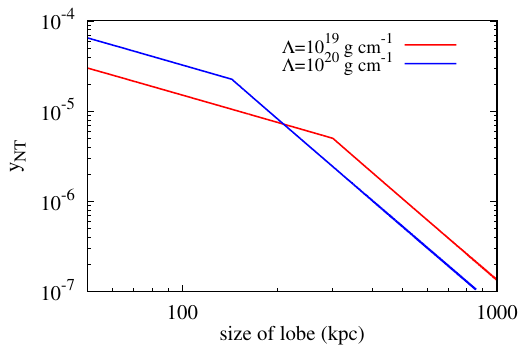}
\caption{
\color{black}$ y_{NT}$ function of the size of the radio lobe for different density of surrounding gas with $\Lambda=10^{19}, 10^{20}$ g cm$^{-1}$ respectively. Jet luminosity $Q_{\rm J}=10^{45}$erg s$^{-1}$, starting redshift $z_{st}=0.5$ and $t_{\rm J}=10^8$ yr.}  
\label{yvsdensity}
\end{figure}

\begin{figure}
\centering
\includegraphics[width=\columnwidth]{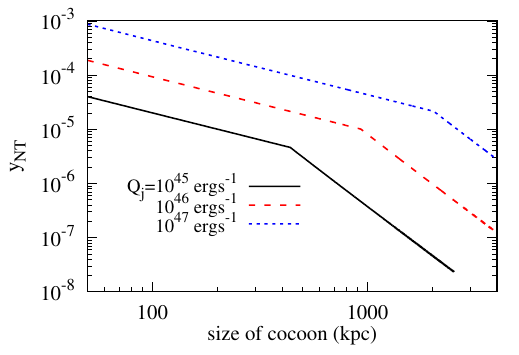}
\caption{The non-thermal CMB distortion amplitude, $y_{\rm{NT}}$, from \color{black}radio lobes \color{black} as a function of their size for different jet luminosities $Q_{\rm J}$.  In each case, the jet starting redshift is $z_{st}\,=\,0.5$ and the jet is evolved upto $z=\,0.3$. The jet lifetime is $t_{\rm J}=10^8 $ yr.}  
\label{yparametervsL3}
\end{figure}

In Fig. \ref{radiopowervsz}, we plot the radio power as a function of size of \color{black}radio lobe \color{black} with different jet starting redshifts and jet luminosities keeping $t_{\rm J}= 10^8$ yr. For a  constant jet luminosity, radio power is independent of the initial redshift. Since we have assumed that the surrounding medium of the \color{black}radio lobe \color{black} has density profiles that is independent of redshift. The fall gets sharper with increasing redshift due to efficient cooling of the electrons at higher redshifts. After the jet stops, as there are no more energetic electrons that emit at 150 MHz, the power falls sharply. With increasing jet luminosity, the radio power increases as expected which is shown in Fig. \ref{radiopowervsz}. \color{black} In Fig. \ref{radiopowervsalpha}, we compare radio power for two different spectral indices. With flatter spectral index, there is significantly higher radio power due to substantial increase of number of electrons at high energy. In Fig. \ref{radiopowervsdensity} and \ref{yvsdensity}, we compare the radio power and $y_{\rm{NT}}$ from radio galaxy lobes for different density of surrounding gas. For higher density, the lobe has a slower growth of size and therefore higher pressure for a chosen jet luminosity. Therefore, $y_{\{rm{NT}}$ is higher for the case of higher density before the stoppage of jet. \color{black}Owing to our assumption of equipartition, a smaller size 
of the radio lobe implies increased number density of electrons, which, in turn,
implies higher magnetic energy density, leading to higher luminosity. \color{black} In Fig. \ref{yparametervsL3}, we show $y_{\rm{NT}}$ as a function of the size of \color{black}radio lobe \color{black} and jet luminosity. The expression for $y_{\rm{NT}}$ is given by
\begin{equation}
y_{\rm{NT}}=\frac{\sigma_{\rm{T}}}{m_{\rm{e}}{c}^2}p_c\times 2L_{\rm J},
\end{equation} 
where the symbols carry their usual meanings. The time taken for light to  travel across the length of a \color{black}radio lobe \color{black} is small compared to the light travel time to reach us. Therefore, the SZ observation of a radio \color{black} lobe \color{black} gives a snapshot of the \color{black}lobe \color{black} during its evolution at the observed redshift. With increase in the size of the \color{black}radio lobe \color{black}, the pressure inside the \color{black}lobe \color{black} falls and, therefore, $y_{\rm{NT}}$ decreases. With increase in jet luminosity, pressure and $y_{\rm{NT}}$ increases. The curves show that in order to achieve $y_{\rm{NT}}\ge 10^{-5}$, either the \color{black}radio lobes \color{black} have to be young (i.e. smaller in size) and the jet has to be active. In other words, it would be difficult to detect non-thermal SZ \color{black} effect \color{black} from `dead' radio galaxies, because their distortion would be $y_{\rm{NT}}\le 10^{-6}$. Also, giant radio galaxies with Mpc size are not favourable. For a jet luminosity of $10^{46}$ erg s$^{-1}$ or higher, radio galaxies with size $\le 200$ kpc are favourable for non-thermal SZ detection. We note that the \color{black}radio lobe \color{black} size depends on the ambient density and would be smaller for a radio galaxy residing in a cluster environment (see also N2010), and may provide with good targets for non-thermal SZ detection. From Fig. \ref{yparametervsL3}, it is interesting to note that a small young radio \color{black}lobe \color{black} of $L_{\rm J}\sim 100-200$ kpc and $Q_{\rm J}=10^{47}$ erg s$^{-1}$ can have  an $y_{\rm{NT}}$ equal in amplitude to thermal SZ from hot ICM of a galaxy cluster. Then taking into account the spectral distortion shape (see Fig. \ref{SZspectrum}), the CMB distortion by a radio \color{black}lobe \color{black} inside a cluster can be a significant part of the SZ distortion from the cluster ICM. Neglecting any $y_{\rm{NT}}$  would result in a source of systematic bias for the SZ measurements towards a cluster. 

\section{Interdependency and degeneracy between parameters of the radio galaxy evolution model}
\label{sec:degeneracy}
\begin{figure}
\centering
\includegraphics[width=\columnwidth]{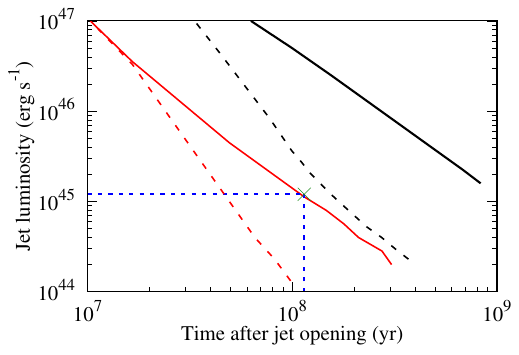}
\caption{Modelling the jet luminosity as a function of jet starting redshifts (or equivalently time elapsed) for two sources selected from Table \ref{tab:table2}. The evolution of the \color{black}radio lobes \color{black} are constrained to match with observations, i.e, to reach a given observed size at a particular redshift of observation. The upper two lines are for the source B2 1358+30C with a \color{black}radio lobe \color{black} size of 1400 kpc  with at $z$ = 0.206 and the lower two lines are for the source 3C 274.1 with a \color{black}radio lobe \color{black} size of 460 kpc at $z$ = 0.422. For each source, the solid and dashed lines are for jet lifetimes of $t_{\rm J}=10^7 \&10^8$ yr.  The green cross shows that a combination of $Q_{\rm J} \sim 10^{45}$ erg s$^{-1}$, $t_{\rm J} = 10^7$ yr and time evolved of $\sim 10^8$ yr would model the radio \color{black} galaxy lobe \color{black} 3C 274.1. }
\label{Ljvszstart}
\end{figure}

In this section, we study the degeneracy between parameters of our radio \color{black} galaxy evolution \color{black} model. Previous discussion shows that the non-thermal SZ \color{black} effect \color{black} depends on the jet luminosity, the starting redshift, and the observed redshift. We scan the parameter space in jet luminosity ($Q_{\rm J}$) and starting redshift of jet ($z_{st}$) which will expand to a particular size at a given redshift $z$. The parameter ($z_{st}$) can be translated to the time interval ($\Delta t$) after jet opening until the observed redshift.  We consider two sources for illustrative purpose: 3C 274.1 and B2 1358+30C (\cite{CMBM2013} , their Table 1), observed at $z=0.422$ and 0.206 respectively. The size of their major axis are 460 kpc and 1400 kpc respectively. We require that the \color{black}radio galaxy lobe \color{black} starting at $z_{st}>z$ with some $Q_{\rm J}$ will grow to size ($2L_{\rm J}$) $460\pm10$ kpc and $1400\pm 10$kpc by  $z=0.422$ and 0.206 respectively. We assume initial size of the \color{black}lobe \color{black} to be 10 kpc which is the size of a galaxy. The degeneracy between the jet luminosity and elapsed time after jet opening formation, for two different jet lifetimes, which satisfies the constraint of getting the observed \color{black}radio galaxy lobe \color{black} size at $z$ is shown in Fig. \ref{Ljvszstart}. The luminosity is chosen to vary between $10^{44}$ erg s$^{-1}$ and $10^{47}$erg s$^{-1}$ \citep{HC2020}. The inferred jet luminosities from observed radio galaxies upto now seem to be below $10^{47}$erg s$^{-1}$. Note that for shorter jet lifetime, luminosity has to be larger so that pressure inside the \color{black}lobe \color{black} is high enough to make it expand to a particular size. For $\Delta t$ less than $10^7$yr, the total energy content in the \color{black}radio galaxy lobe \color{black} is $Q_{\rm J}\Delta t$ irrespective of jet lifetime being $10^7$ or $10^8$ yr and, therefore, the curves with the two different lifetimes merge. 

\begin{figure}
\centering
\includegraphics[width=\columnwidth]{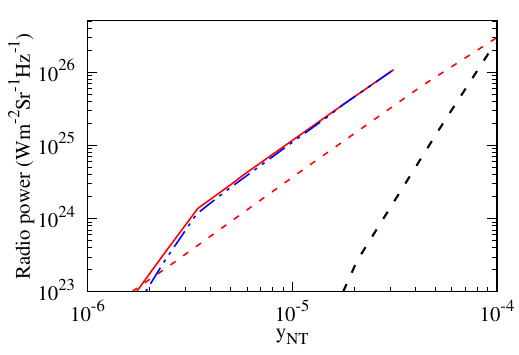}
\caption{$y_{\rm{NT}}$ vs radio power at 150 MHz for two sources selected from Table \ref{tab:table2}. The evolution of the \color{black}radio lobes \color{black} are constrained to match with observations, i.e, to reach a given observed size at a particular redshift of observation. The dashed lines are for the source 3C 274.1 having \color{black}radio lobe \color{black} size of 460 kpc at $z\,=\,0.422$, with the black thick dashed line for jet lifetime $t_{\rm J}\,=\,10^7$yr and thin red dashed line for $t_{\rm J}\,=\,10^8$yr. The other source, B2 1358+30C, having size of 1400 kpc at $z\,=\,0.206$, is shown with the red solid line. Note, that B2 1358+30C would have been almost degenerate in $y_{\rm{NT}}$-radio power plane with a source of size 1400 kpc at $z\,=\,0.6$ \& $t_{\rm J}=10^8$ yr, shown with the blue dot-dashed line. }  
\label{yvsLr}
\end{figure}

Next, we plot the degeneracies between the estimated $y_{\rm NT}$ and the radio power at 150 MHz in Fig. \ref{yvsLr}  for the lifetimes $10^7$yr and $10^8$yr. We take the same two sources depicted in Fig. \ref{Ljvszstart}. We also plot the scenario in which the radio \color{black}lobe \color{black} of size 1400 kpc is observed at $z$=0.6 instead $z$=0.206  and show how it is almost degenerate with the other radio \color{black}lobe \color{black} of size 1400 kpc, but at $z$=0.206.  We have assumed \color{black} electron momentum spectral index to be $\alpha_p=3$ \color{black} for the rest of the calculations as the detected radio galaxies used in this work have observed \color{black} momentum \color{black} spectral index $\sim 3$. Compared to $\alpha_p=2.3$, this results in a reduction of the radio power by more than an order of magnitude as electron population dies off steeply.  For the range of luminosities chosen, the radio power drops drastically before the size of the \color{black}lobe \color{black} reaches $\sim$ 1000 kpc for $t_{\rm J}=10^7$ yr as the jet dies off. Radio power is directly correlated with $y_{\rm{NT}}$ as the synchrotron and SZ signal comes from same population of high energy electrons \color{black} assuming constant spectral index and equipartition\color{black}. Higher $y_{\rm{NT}}$ corresponds to higher $Q_{\rm J}$, and consequently, higher radio power, as more number of high energy particles are emitted by the jet. 

It is interesting to note that although the pressure inside a \color{black}radio lobe \color{black} changes as the \color{black}lobe \color{black} expands,  for a particular \color{black}lobe \color{black} size at any instant,  both the radio power and $y_{\rm{NT}}$ are independent of the redshift. This may be understood from Eq. \ref{cocoonsize}. Given a particular size, since we assume the density of the surrounding medium to be independent of redshift, the pressure inside the \color{black}radio galaxy lobe \color{black} and, therefore $y_{\rm{NT}}$, is independent for all redshifts. Radio power is the same once the jet lifetime is held constant and the injected particles by the jet becomes constant for all cases. For a fixed radio power, $y_{\rm{NT}}$ increases with decreasing \color{black}lobe \color{black} size as pressure increases with decreasing volume.  For $10^7$yr$<\Delta t<10^8$yr, the jet with lower lifetime stops supplying particles, which then leads to reduction in emitted radio power for the jet with $t_{\rm J}$=$10^7$yr compared to one with $t_{\rm J}$=$10^8$ yr. The size of the \color{black}radio lobe \color{black} also makes a difference. For a given jet luminosity, the \color{black}lobe\color{black} can expand to 460 kpc relatively easily compared to 1400 kpc for $t_{\rm J}=10^7$ yr while the jet is still on. Therefore, a \color{black}radio lobe \color{black} of size 460 kpc has significant radio power for both lifetimes, $t_{\rm J}$=$10^7$ and $10^8$ yr. In contrast, for a 1400 kpc sized \color{black}radio lobe \color{black}, there is practically no radio power for the lower lifetime of $t_{\rm J}$=$10^7$ yr . We conclude that selecting \color{black}radio galaxy lobes \color{black} of high radio power and small size (which is more likely in cluster environment, as discussed earlier) would pay off towards detection of non-thermal SZ \color{black}signal\color{black}. 

\begin{figure}
\centering
\includegraphics[width=\columnwidth]{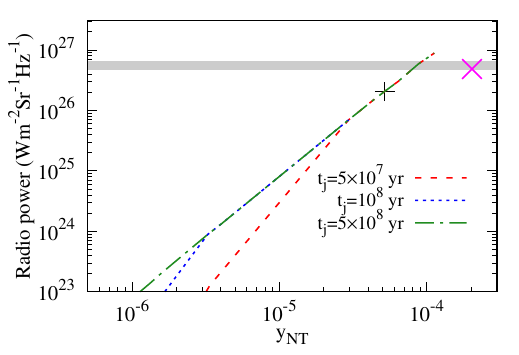}
\caption{Similar to Fig. \ref{yvsLr}, $y_{NT}$ vs radio power at 150 MHz for radio \color{black}lobes \color{black} constrained to have size $2L_{\rm J}\,=\,1$ Mpc at $z_{\rm obs}\,=\,0.3$ (to model the source observed by \citet{MDCMSNW2017}) for different jet lifetimes. The jet luminosity is varied between $10^{44}-10^{48}$ erg s$^{-1}$.  The grey horizontal band shows the radio power observed by \citet{MDCMSNW2017}. The black cross corresponds to the radio luminosity and $y_{\rm{NT}}$ which can be achieved for a 1 Mpc \color{black}lobe \color{black} with a $Q_{\rm J}$ of $10^{47}$ erg s$^{-1}$. In contrast, for the \color{black}radio lobe \color{black} being closer to us (i.e., having a lower $z$, see text for details) such that the size is 200 kpc, but with the same $Q_{\rm J}$,  the corresponding point is shown by the magenta cross which touches the grey band.}  
\label{yvsLrvstj}
\end{figure}

It is interesting to consider the recent detection of non-thermal SZ \color{black}signal \color{black} from a radio galaxy by \cite{MDCMSNW2017} \color{black} assuming it to be near the bullet cluster at $z\approx 0.3$. With this assumption\color{black}, the distance of the hotspot from the radio galaxy core is estimated to be $\approx 1$ Mpc. The radio power at $5.5$ GHz is $4 \times 10^{28}$ W, over a $2$ GHz bandwidth (S. Malu, 2020, private communication). Considering the observed bandwidth, this implies radio power of $2 \times 10^{19}$ W Hz$^{-1}$. Considering the angular area of the lobe to be $4' \times 4'\approx 1.35 \times 10^{-6}$ sr, and 
assuming the minor axis to be half the hotspot distance, as in our model, the radio power turns out to be $3 \times 10^{25}$ W Hz$^{-1}$ sr$^{-1}$.
Then, with a spectral index of $-1$ (\color{black} corresponding to $\alpha_p=3$\color{black}), this implies a power at $150$ MHz $\approx 5.5 \times 10^{26}$ W Hz$^{-1}$sr$^{-1}$. This level of radio power is shown with 15 percent measurement uncertainty with a grey horizontal band in Fig. \ref{yvsLrvstj}. In this figure, we also show the radio power as a function of $y_{\rm{NT}}$ for different jet lifetimes $t_{\rm J}$. The jet luminosity is varied between $10^{44}-10^{48}$ erg s$^{-1}$. In their paper, \cite{MDCMSNW2017} reported an observed non-thermal SZ distortion with $y=2\times 10^{-5}$. \color{black} Note that in our calculation, the jet luminosity has to be slightly higher than $10^{47}$ erg s$^{-1}$ to explain the observed radio power for a 1 Mpc \color{black}radio lobe \color{black}. The radio power for a 1 Mpc \color{black}lobe \color{black} which can be achieved by jet luminosity $10^{47}$ erg s$^{-1}$ is shown as the black cross in Fig. \ref{yvsLrvstj}. The required jet luminosity to explain the radio power turns out to be $\sim 3\times 10^{47}$ erg s$^{-1}$. \color{black} Alternatively, the radio galaxy may be a foreground object, in which case its size is smaller, and a smaller jet luminosity is needed to explain this observation. \color{black} In order to \color{black} determine the requirements of energetics, we fix the jet luminosity at $Q_{\rm J}=10^{47}$ erg s$^{-1}$ and vary the size of the \color{black}lobe \color{black}. We find that for a \color{black}radio lobe \color{black} with size 200 kpc, the radio power can be explained by $Q_{\rm J}=10^{47}$ erg s$^{-1}$ (shown with the magenta cross in the same figure) \color{black}. This foreground object has to be, then, located at $z=0.05$ to have the observed angular size in the sky.

We would like to point out that \cite{MDCMSNW2017}  assumed non-relativistic distortion in obtaining their value of $y\,=\, 2\times10^{-5}$ . However, one actually measures  $\Delta I_{\nu}$  which can be written as,
\begin{eqnarray}
\Delta I_{\nu}(x)&=&yg_{\rm{T}}(x) \;\;\;\;\;\;\;\;\;\;\; {\rm (Malu\,et. al.\,(2017))} \nonumber \\
&=&y_{\rm{NT}}g_{\rm{NT}}(x)\;\;\; {\rm (this \, work)} \;,
\label{intdistortion}
\end{eqnarray}
where $y$ and $g_{\rm{T}}(x)$ is the non-relativistic $y$-distortion and the spectrum respectively, and similarly $y_{\rm{NT}}$ and $g_{\rm{NT}}(x)$ for non-thermal distortions. Therefore, there is a degeneracy between  $y_{\rm{NT}}$ and $g_{\rm{NT}}(x)$ (or $p_{\rm{min}}$) for a given $\Delta I_{\nu}(x)$. This is true for the measurement of $\Delta I_{\nu}(x)$ at a single frequency which was the case for the reported detection at 18 GHz. The magnitude of $g_{NT}(x)$ is lower than $g_{\rm{T}}(x)$ for all $x$ (see Fig. 6 of \citep{EK2000}) and this difference depends sensitively on the value of $p_{\rm{min}}$. At $x=0.35$ (corresponding to frequency 18 GHz), the value of $g_{\rm{T}}=0.24$ while $g_{\rm{NT}}=0.1$ and 0.03 for $p_{\rm{min}}=1$ and 2 respectively (as seen in Fig. \ref{gx}). Therefore, for $p_{\rm{min}}=1$ and 2, $y_{\rm{NT}}$ is higher than $y=2\times 10^{-5}$ by a factor of 2.5 and 10 respectively, and can easily reach $y_{\rm{NT}}=2\times 10^{-4}$ for $p_{\rm{min}}=2$. 

\begin{figure}
\centering
\includegraphics[width=\columnwidth]{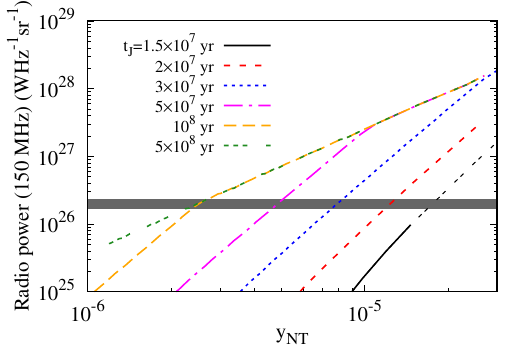}
\caption{\color{black}Same calculations as in Fig. \ref{yvsLrvstj} but with $\alpha_p=2.3$. The jet luminosity is varied between $10^{44}-10^{47}$ erg s$^{-1}$ except for the thin dashed black line for which the highest jet luminosity is chosen to be $10^{48}$ erg s$^{-1}$\color{black}}  
\label{yvsLrvstjalpha=2.3}
\end{figure}

The degeneracy between $y_{\rm{NT}}$ and $p_{\rm{min}}$ was already noted in \cite{MDCMSNW2017} (their Fig. 2). They obtain an upper limit on $p_{\rm{min}}$ between 5 to 10 using X-ray constraints. Note, that their $y$-value is for non-relativistic SZ where the electron spectrum ($p_{\rm{min}}$) does not enter. Moreover, without a radio \color{black} galaxy \color{black} evolution model, they could not relate $y$-value to the size of the \color{black}radio galaxy lobe \color{black}. In contrast, our approach of using a detailed radio \color{black}galaxy \color{black} evolution model leads us to predict a value of $y_{\rm{NT}}$ for a given size of the \color{black}radio lobe \color{black} and radio power. From our model, $y_{\rm{NT}}$ for 1 Mpc object turns out to be $6-7\times 10^{-5}$ (\color{black} corresponding to jet luminosity of $\sim 3\times 10^{47}$ erg s$^{-1}$). This value of $y_{\rm{NT}}$ is consistent with the the value of $p_{\rm{min}}=1$. \color{black} Even if we mistake the source to be a foreground object of size 200 kpc, \color{black} the obtained value of $y_{\rm{NT}}$ from our evolutionary model is $\sim 2\times 10^{-4}$ which is consistent with $p_{\rm{min}}=2$. The increase in $y_{\rm{NT}}$ for 200 kpc \color{black}lobe \color{black} as compared to 1 Mpc size \color{black}lobe \color{black} necessarily leads to increase in $p_{\rm{min}}$ (or decrease in value of $g_{\rm{NT}}(x)$ at $x=0.35$) such that $\Delta I_{\nu}(x)$ is unchanged as in Eq. \ref{intdistortion}.  \color{black}  \color{black}Note, that irrespective of the degeneracy between $y_{\rm{NT}}$ and $p_{\rm{min}}$, an arbitrary increase in $y_{\rm{NT}}$ in our radio galaxy evolutionary model would require high non-thermal pressure which necessarily requires the size of the radio lobe to be unrealistically small for a given radio luminosity. \color{black} Therefore, the observed radio power and size of the \color{black}radio lobe \color{black} (assuming the \color{black}lobe \color{black} to be at $z\sim0.3$) is consistent with \color{black}$y_{NT}=6-7\times10^{-5}$ which requires $p_{\rm{min}}\sim1$ \color{black}. This is the first estimate of the lower energy cutoff of non-thermal electron population in radio galaxy \color{black}lobe \color{black} using SZ effect. \color{black}
In Fig. \ref{yvsLrvstjalpha=2.3}, we repeat the above calculation but with a flatter spectral index $\alpha_p=2.3$. We find that for $t_{\rm J}\sim 10^7$ yr, our model predicts $y_{\rm{NT}}\sim 2\times 10^{-5}$ but with jet luminosity higher than $10^{47}$ erg s$^{-1}$. But to explain \cite{MDCMSNW2017} result, $y_{\rm{NT}}$ has to be $\sim 10^{-4}$ as $g_{\rm{NT}}(x)$ at $x=0.35$ is $\sim 0.05$ for the spectral index chosen. Therefore, our galaxy evolution model is inconsistent with a spectral index of 2.3. We also note that this object is at a sufficiently low redshift, for which our assumption of the ambient density is reasonable. \color{black}
\par
If we have measurements of the CMB distortions at multiple frequencies, we can directly measure the value of $p_{\rm{min}}$ from the shape of distortion and break the degeneracy between $y_{\rm{NT}}$ and $g_{\rm{NT}}(x)$. The viability of such measurements with upcoming CMB experiments is discussed in Sec. 6.


\subsection{Comparison with Colafransesco et. al.}
We compare our non-thermal SZ calculation with the radio galaxies listed in  \cite{CMBM2013}. The authors assumed a static electron distribution where the electron number density is given by,
\begin{equation}
N_{\rm{e}} (p,r)=k_0g_{\rm{e}}(r)A(p_1,p_2,\alpha_p)p^{-\alpha_p},
\end{equation}
The value of $k_0$ is fixed to be 2.6 $\rm{cm^{-3}}$ \color{black} and the average value of observed $\alpha_p$=3 \color{black}  which is obtained from fitting data to radio and X-ray observations. The electron number density is assumed to be constant, so that $g_{\rm{e}}(r)=1$. $A$ is a normalizing constant such that $\int_{p_1}^{p_2} Ap^{-\alpha_p}=1$. $p_1$ is fixed to be 1. This can be converted to cutoff in $\gamma$ as $\gamma=\sqrt{(1+p_1^2)}$. The size of galaxy is assumed to be ellipsoidal with major and minor axis as given in Table 1. The line of sight is assumed to be the major axis. From these information one can calculate the optical depth for CMB photons.

In Table \ref{tab:table2}, we list the sources, their size, radio power at 150 GHz which were considered in \cite{CMBM2013}. We then predict the $y_{\rm{NT}}$, an estimate of $t_{\rm J}$ and flux density for Simons Observatory at two frequencies, which are allowed by our model and satisfy the size and radio flux density as given. The beam and instrument noise for Simons Observatory are listed in Table 1 of \cite{SO2019}. We use the best case scenario with specifications for LAT ($f_{sky}=0.4$) with FWHM as the beam. The expected noise are 6.3 and 37  $\mu$K-arcmin for the two frequencies in Table \ref{tab:table2}. The $t_{\rm J}$ values listed in Table \ref{tab:table2}  should be considered as an lower limit of jet lifetime that satisfies the constraints. \color{black} We see that for most of the sources, the required jet luminosity needs to be higher than $10^{47}$ erg s$^{-1}$. Therefore, we have allowed the jet luminosity  to vary between $10^{44}$ and $10^{48}$ erg s$^{-1}$. \color{black} For a given radio galaxy size, there is a three-way degeneracy between the jet luminosity, the lifetime of jet and the starting redshift of jet ($z_{st}$) or the time-interval between jet starting redshift and observed redshift ($\Delta t$) (Fig. \ref{Ljvszstart}). For a given jet lifetime and size, we can increase $\Delta t$ by reducing jet luminosity or vice versa. Therefore, a bound on jet luminosity gives a bound on $\Delta t$. The radio power from a galaxy drops sharply once jet is off i.e. we should not observe radio flux density if $\Delta t>t_{\rm J}$. This criteria and prior condition on jet luminosity gives a lower bound on $t_{\rm J}$. However, radio observation does not give an upper bound on $t_{\rm J}$ since the radio \color{black}lobes \color{black} becomes invisible in radio as it grows too big and faint . In contrast, the radio galaxy should be observable in SZ even after jet goes off. Therefore, SZ observations can put an upper bound on jet lifetimes of these radio galaxies. \color{black} In Table \ref{tab:table3}, we show the predicted $y_{\rm{NT}}$ and lower estimate of $t_{\rm J}$ assuming the spectral index of electron distribution to be 2.3. For this case, the predicted $y_{\rm{NT}}$ are orders of magnitude lesser compared to the case when the index is 3. This can be understood from the discussion of Fig. \ref{radiopowervsalpha}. For same jet luminosity, the radio power from an electron distribution with flatter spectral index is higher compared to a steeper spectrum. Therefore, to explain a given radio power, the inferred jet luminosity is smaller in case of flatter spectral index and hence smaller pressure inside the lobe and smaller $y_{\rm{NT}}$.
   \begin{table*}
   \begin{tabular}{l|c|c|c|c|c|c|c|r} 
    source & $z$ & angular size & size & Flux & $y_{NT}$ & $t_{\rm J}$ & flux ($\mu$K-arcmin) &  flux ($\mu$K-arcmin) \\
           &    & arcsec (major axis)  &   kpc      &     mJy   &          &        yr   & (150 GHz) & (280 GHz) \\
    \hline
    CGCG 186-048 & 0.063 & 388 & 485 & 1 & $2\times 10^{-5}$ & $5\times 10^7$ & 399 (63$\sigma$) & 6.1 (1.6$\sigma$) \\
    B2 1158+35 & 0.55 & 70 & 462 & 1.3 & $2\times 10^{-4}$ & $5\times 10^7$ & 130 (21$\sigma$) & 20 (0.5$\sigma$)\\
    3C 270 & 0.0075 & 577 & 93 & 113 & $1\times 10^{-5}$ & $  10^7$ & 441 (70$\sigma$) & 68 (1.8 $\sigma$)\\
     87GB 121815.5+.. & 0.2 & 924 & 3141 & 0.43 & $1\times 10^{-5}$ & $ 10^8$ & 1131 (180$\sigma$) & 174 (4.7$\sigma$) \\
     3C 274.1 & 0.422 & 89 & 508 & 30 & $6\times 10^{-4}$ & $5\times 10^7$ & 630 (100$\sigma$) & 174 (2.6$\sigma$) \\
     4C +69.15 & 0.106 & 822 & 1646 & 26 & $1\times 10^{-4}$ & $5\times 10^7$ & 8956  (1421$\sigma$) & 1378 (37$\sigma$)\\
     3C 292 & 0.71 & 64 & 473 & 16 & $8\times 10^{-4}$ & $ 5\times 10^7$ & 434 (69$\sigma$) & 67 (1.8$\sigma$)\\
     B2 1358+30C & 0.206 & 408 & 1421 & 0.28 & $2\times 10^{-5}$ & $ 10^8$ & 441 (70$\sigma$) & 68 (1.8$\sigma$) \\
     3C 294 & 1.779 & 29 & 253 & 2 & $4\times 10^{-4}$ & $ 10^7$ & 45 (7$\sigma$) & 7 (0.2 $\sigma$) \\
     PKS 1514+00 & 0.052 & 519 & 543 & 480 & $5\times 10^{-4}$ & $ 5\times 10^7$ & 17851  (2833$\sigma$) & 2746 (74$\sigma$)\\
     GB1 1519+512 & 0.37 & 312 & 1646 & 22 & $3\times 10^{-4}$ & $ 10^8$ & 3871 (614$\sigma$) & 595 (16$\sigma$)\\
      3C 326 & 0.0895 & 684 & 1177 & 22 & $ 10^{-4}$ & $ 10^8$ & 6201 (984$\sigma$) & 954 (26$\sigma$)\\
     7C 1602+3739 & 0.814 & 100 & 778 & 0.33 & $ 10^{-4}$ & $ 5\times 10^7$ & 132 (21$\sigma$) & 20 (0.5$\sigma$)\\
     MRK 1498 & 0.0547 & 583 & 639 & 15 & $2\times 10^{-4}$ & $ 5\times 10^7$ & 9010 (1430$\sigma$) & 1386 (37$\sigma$)\\
      B3 1636+418 & 0.867 & 57 & 452 & 16 & $ 10^{-4}$ & $ 5\times 10^7$ & 43 (6.8$\sigma$) & 7 (0.2$\sigma$)\\
      Hercules A & 0.154 & 200 & 551 & 253 & $9\times 10^{-4}$ & $ 5\times 10^{7}$ & 4772 (750$\sigma$) & 734 (20$\sigma$)\\
      B3 1701+423 & 0.476 & 120 & 735 & 1.1 & $ 10^{-4}$ & $ 5\times 10^7$ & 191 (30$\sigma$) & 29 (0.8$\sigma$)\\
      4C 34.47 & 0.206 & 92 & 320 & 9.5 & $3\times 10^{-4}$ & $ 5\times 10^{7}$ & 337 (53$\sigma$) & 52 (1.5$\sigma$)\\
      87GB 183438.3+.. & 0.5194 & 69 & 443 & 3.9 & $3\times 10^{-4}$ & $ 5\times 10^{7}$ & 190 (30$\sigma$) & 29 (0.8$\sigma$)\\
      4C +74.26 & 0.104 & 773 & 1522 & 94 & $3\times 10^{-4}$ & $ 5\times 10^7$ & 23760 (3771$\sigma$) & 3655 (99$\sigma$)\\
      RG01 \citep{MDCMSNW2017} & 0.3 & 240 & 1100 & $7\times 10^{-5}$ & 2700 & $10^8$ & 534 (85$\sigma$) & 82.2 (2.2$\sigma$) \\
      
   \end{tabular}
  \caption{We list the redshifts of the sources, size (major axis), radio flux density at 150 GHz given in Table 1 and 2 of \citep{CMBM2013}. We predict the corresponding $y_{\rm{NT}}$, $t_{\rm J}$ and flux density for the source with Simons Observatory \citep{SO2019} or CMB-S4 \citep{S42019} type experiment to match the observation of size and radio power, allowed by our model for radio galaxy. We have used the best case noise limit (LAT, $f_{sky}=0.4$) of Simons Observatory to derive the significance of the detection. For non-thermal spectrum, we choose $p_{\rm{min}}=1$. We have abbreviated a couple of source names to make space. \color{black} Please note that for radio galaxy RG01, the flux density is given at 5.5 GHz. We also assume the source to be at $z$=0.3 (at the location of bullet cluster). If the source is a foreground object, then the physical properties of the galaxy will be different (see text).\color{black} }
    \label{tab:table2}
\end{table*}

 \begin{table*}
   \begin{tabular}{l|c|r} 
    source & $y_{NT}$ & $t_{\rm J}$ \\
     &  &  yr \\
    \hline
    CGCG 186-048 & $3\times 10^{-7}$ & $3\times 10^8$ \\
    B2 1158+35 & $2\times 10^{-6}$ & $2\times 10^8$ \\
    3C 270  & $6\times 10^{-7}$ & $  10^8$ \\
     87GB 121815.5+.. & $3\times 10^{-7}$ & $ 5\times 10^8$ \\
     3C 274.1 & $8\times 10^{-6}$ & $ 10^8$ \\
     4C +69.15 & $2\times 10^{-6}$ & $ 10^8$ \\
     3C 292 & $7\times 10^{-6}$ & $ 10^8$ \\
     B2 1358+30C & $3\times 10^{-7}$ & $ 5\times 10^8$ \\
     3C 294 & $9\times 10^{-7}$ & $ 5\times 10^7$ \\
     PKS 1514+00 & $4\times 10^{-6}$ & $ 10^8$ \\
     GB1 1519+512 & $5\times 10^{-6}$ & $ 10^8$ \\
      3C 326 & $ 10^{-6}$ & $ 2\times 10^8$ \\
     7C 1602+3739 & $ 10^{-6}$ & $ 3\times 10^8$ \\
     MRK 1498 & $6\times 10^{-7}$ & $ 3\times 10^8$ \\
      B3 1636+418 & $ 10^{-6}$ & $ 3\times 10^8$ \\
      Hercules A & $9\times 10^{-6}$ & $ 10^{8}$ \\
      B3 1701+423 & $ 9\times 10^{-7}$ & $ 2\times 10^8$ \\
      4C 34.47 & $2\times 10^{-6}$ & $ 10^{8}$ \\
      87GB 183438.3+.. & $3\times 10^{-6}$ & $10^{8}$ \\
      4C +74.26 & $2\times 10^{-6}$ & $ 2\times 10^8$ \\

   \end{tabular}
  \caption{\color{black}Same as Table \ref{tab:table2} but with a flatter spectral index $\alpha_p=2.3$. We only show the predicted $y_{\rm{NT}}$ and estimated lower estimate of $t_{\rm J}$.\color{black} }
    \label{tab:table3}
\end{table*}

\section{Detection prospects of non-thermal SZ with future CMB experiments}
\label{sec:detection}
In this section, we study the feasibility of detecting non-thermal SZ with future CMB experiments. The distortion in intensity $I_{\nu}$ of CMB at a frequency $\nu$ and towards a localized concentration of hot electrons can be written as,
\begin{equation}
\Delta I_{\nu}=\Delta I_{\rm{y}}+\Delta I_{\rm{\mu}}+\Delta I_{\rm{kSZ}}+\Delta I_{\rm{rSZ}}+\Delta I_{\rm{NT}},
\end{equation}
where $\Delta I_{\rm{y}}$ is the typically dominant non-relativistic y distortion, $\Delta I_{\rm{\mu}}$ is the $\mu$ distortion, $\Delta I_{\rm{kSZ}}$ is the temperature shift of the CMB black body due to kSZ effect \citep{SZ1980}, $\Delta I_{\rm{rSZ}}$ is the relativistic SZ signal from a massive galaxy cluster and $\Delta I_{\rm{NT}}$ is the non-thermal SZ distortion. The expression for $\Delta I_{\rm{y}}$, $\Delta I_{\rm{\mu}}$ and $\Delta I_{\rm{kSZ}}$ are given by \citep{ZS1969,Sz19701,Is19751},
\begin{equation}
\Delta I_{\rm{y}}=y\frac{2\rm{h}{\nu}^3}{\rm{c^2}}\frac{x\rm{e^x}}{(\rm{e^x}-1)^2}\left[x\frac{\rm{e^x}+1}{\rm{e^x}-1}-4\right]
\end{equation}
\begin{equation}
\Delta I_{\rm{\mu}}=\mu\frac{2\rm{h}{\nu}^3}{\rm{c^2}}\frac{\rm{e^x}}{(\rm{e^x}-1)^2}\left[\frac{x}{2.19}-1\right]
\end{equation}
\begin{equation}
\Delta I_{\rm{kSZ}}=\color{black}\left(\frac{\Delta T}{T}\right)\color{black}\frac{2\rm{h}{\nu}^3}{\rm{c^2}}\frac{\rm{e^x}}{(\rm{e^x}-1)^2}
\end{equation}
The $y$-distortion can have contribution from both pre-recombination universe as well as post-recombination universe while $\mu$-distortion can only be created at redshifts greater than $2\times 10^5$. 
 The SZ spectrum from electrons with energy $\gtrsim$ keV is obtained by including Klein-Nishina corrections for Compton scattering which can be calculated perturbatively as an expansion in $\frac{T_{\rm{e}}}{m_{\rm{e}}}$ \citep{IKN1998,CL1998,SS1998,DHPS2001}. For example, The temperature of massive galaxy cluster with mass $10^{14}$ times solar mass can be $\sim$ 5 keV \citep{EBCB2018}, and with a typical $y \sim 10^{-5}$.
\begin{figure}

\includegraphics[width=\columnwidth]{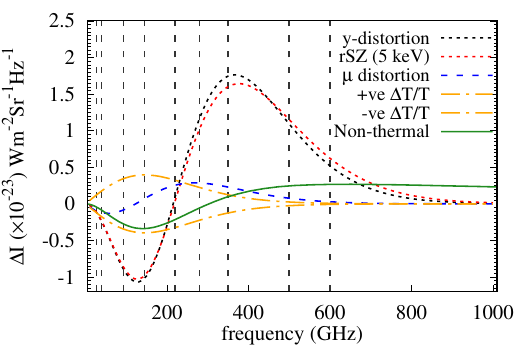}
\caption{The frequency dependence of various possible distortions from CMB black body including the non-thermal SZ distortion (green solid line).  The amplitude of distortion is chosen to be $10^{-6}$ for each type. Moreover, for the non-thermal spectrum, we choose $p_{\rm{min}}=1$. The vertical dashed grey lines correspond to current and upcoming CMB observational frequencies.} 
\label{SZspectrum}
\end{figure}
\par

The unique spectral shapes of the different SZ distortions can be used to separate them, hence isolating the non-thermal SZ from other dominant CMB fluctuations, at relevant angular scales. In Fig. \ref{SZspectrum}, we plot the intensity of different distortions as discussed above. The spectral shapes of these distortions are different from each other. Therefore, with a multifrequency study, we can distinguish non-thermal distortion from other forms of distortions. To check if non-thermal distortion can be mimicked by combination of other distortions, we find the best fit to non-thermal distortion shape from a linear combination of $y$, $\mu$, $\rm{kSZ}$ and rSZ distortions i.e. we want to write  $\Delta I_{NT}$ as,
\begin{equation}
 \Delta I_{\rm{NT}}=y\Delta I_{\rm{y}}+\mu \Delta I_{\rm{\mu}}+\color{black}\left(\frac{\Delta T}{T}\right)\color{black}\Delta I_{\rm{kSZ}}+y_{\rm{rSZ}}\Delta I_{\rm{rSZ}},
 \end{equation}
 where $y, \mu, \color{black}\frac{\Delta T}{T}\color{black}$ and $y_{\rm{rSZ}}$ are the best fit parameters. The basic idea is that if the non-thermal distortions cannot be written as a linear combination of other forms of distortions i.e. if there is a non-zero residual after the best fit has been removed from total distortion, then we can detect the non-thermal part of the total distortion as the residuals of total distortion signal.

\par  
\begin{figure}

\includegraphics[width=\columnwidth]{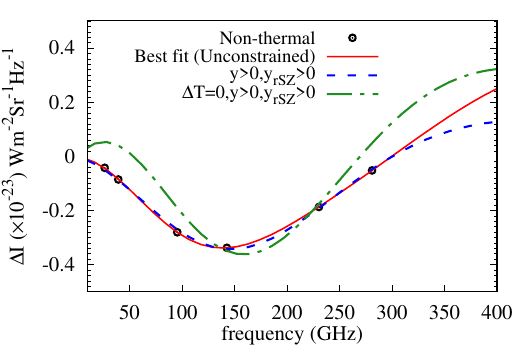}
\caption{ The intensity of non-thermal distortions at frequencies 27, 39, 93, 145, 225 and 280 GHz as proposed for the upcoming Simons Observatory (shown with open circles), having the amplitude $y_{\rm{NT}}\,=\,10^{-6}$.  The best fit to non-thermal distortion using a linear combination of $y$, $\mu$, $\rm{kSZ}$ and rSZ distortions is shown by the solid red line, which perfectly passes through the observed points. The unconstrained fit is shown by red-solid line whereas constraints on the amplitude  of thermal SZ results in the blue dashed line and an additional constraint of no kinematic SZ gives the green dot-dashed line.} 
\label{Simonsfit}
\end{figure}

\par
In Fig. \ref{Simonsfit}, we plot the best fit with six frequency channels of Simons Observatory which will start observation in near future. For the unconstrained best fit, the best fit parameter of $y$, $y_{\rm{rSZ}}$, $\mu$ and $\Delta T/T$ can have positive and negative sign. With these unconstrained parameters, we have a very good fit to non-thermal distortion. Note that even in the absence of a galaxy cluster, the $y$ and $\mu$-type distortion can be  created in pre-recombination universe from acoustic damping, baryon cooling or injection of electromagnetic energy from decay or annihilation of unstable particles etc. \citep{CS2012,KSC2012,CKS2012}. While acoustic damping and injection of energy can give rise to positive $y$ or $\mu$-type distortion, baryon cooling gives rise to negative $\mu$ distortion of the order $10^{-8}-10^{-9}$.  In post recombination universe, the hotter electrons in the structures boost CMB photons to higher energy. Therefore, it is reasonable to assume that $y$ and $y_{\rm{rSZ}}$ should be positive while the $\rm{kSZ}$ temperature shift can be positive and negative as this distortion is produced by moving electrons. With the constraints that $y$ and $y_{\rm{rSZ}}$>0, we can still manage to have a very good fit to non-thermal spectrum in the frequency range 20-300 GHz. Note that, a negative $\rm{kSZ}$ temperature shift can mimic non-thermal SZ spectrum by shifting the null point to higher frequencies. As can be seen from Fig. \ref{SZspectrum}, the shape of a negative $\rm{kSZ}$ is roughly similar to non-thermal distortion for frequency $\lesssim$ 300 GHz. Once, we ignore $\rm{kSZ}$ with the positivity condition on $y$ and  $y_{\rm{rSZ}}$, we are able to distinguish non-thermal SZ signal as there are non-zero residual at some frequencies (Fig. \ref{residual2}). For an order of magnitude estimate of temperature shift due to $\rm{kSZ}$, we consider the source Hercules A from Table \ref{tab:table2} with highest $y_{\rm{NT}}$ parameter ($y_{\rm{NT}}=9\times 10^{-6}$). The optical depth of this source turns out to be $\sim 10^{-4}$. With $v/c\sim 10^{-3}$, we see that the distortion due to $\rm{kSZ}$ are of the order $\frac{v}{c}\tau\sim 10^{-7}-10^{-6}$. \color{black} In the above estimate, we have assumed that the positive and negative kSZ contribution from the jet would cancel and only the virial velocity of cluster will contribute to the kSZ signal. For a spherical source, there is full cancellation. For non-spherical source, there can be a small effect due to incomplete cancellation. However, the more the inclination, the kSZ is also low because the component of radial velocity is smaller (see \cite{ADPG1996} and \cite{MNC2001} as example). \color{black} Therefore, we can safely ignore the $\rm{kSZ}$ temperature shift and consequently will be able to differentiate non-thermal SZ from other distortions, for this source.

\begin{figure}
\includegraphics[width=\columnwidth]{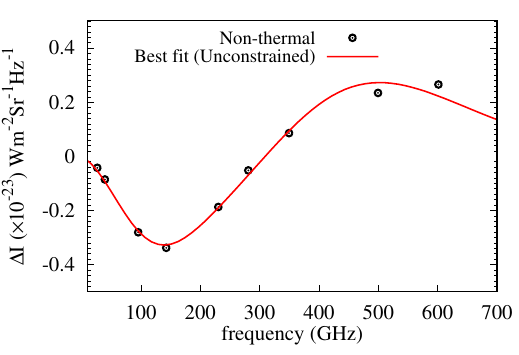}
\caption{Same as Fig. \ref{Simonsfit}, but for a hypothetical experiment with the six frequencies of Simons observatory and additional frequency bands at 350 GHz, 500 GHz and 600 GHz. A linear combination of $y$, $\mu$, $\rm{kSZ}$ and rSZ distortions can no more pass through all the observed points.} 
\label{hypotheticalfit}
\end{figure}
\begin{figure}
\includegraphics[width=\columnwidth]{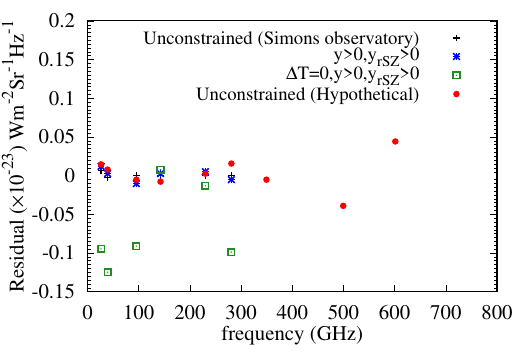}
\caption{Residual after subtracting best fit linear combination of $y$, $\mu$, $\rm{kSZ}$ and rSZ distortions from non-thermal distortions corresponding to Figs. \ref{Simonsfit} \& \ref{hypotheticalfit}.  } 
\label{residual2}
\end{figure}
\par
For a completely unambiguous detection without any assumption on the $y$, $\mu$, $\rm{k}$SZ and rSZ distortions, it is clear from the above discussion that the location of the  null point of the total distortion may not be a reliable signature of non-thermal SZ from future CMB experiments. However, as can be seen from Fig. \ref{SZspectrum}, the intensity of  $y$, $\mu$, $\rm{kSZ}$ and rSZ distortions after 400 GHz starts to decrease in magnitude and approach zero while for non-thermal SZ, the spectrum is relatively flat. Therefore, high frequency channels ($>400 \rm{GHz}$) would be able to differentiate non-thermal distortion from others. We design a hypothetical experiment with all the frequency band of Simons Observatory and add three more frequency channel at 350 GHz, 500 GHz and 600 GHz. In this setup, we are able to distinguish non-thermal distortion from others without putting any constraints on best fit parameters of  $y$, $\mu$, temperature shift due to $\rm{kSZ}$ and rSZ distortions as can be clearly seen in Fig. \ref{hypotheticalfit} and Fig. \ref{residual2}.  
\par
  Some of the sources listed in Table \ref{tab:table2} can be very good candidates for Simons Observatory or CMB-S4 experiments due to high signal to noise ratio at 150 and 280 GHz frequency bands. However, for unambiguous detection, we will need higher frequency bands. Experiments like Probe of Inflation and Cosmic Origins (PICO) \citep{PICO2019} with many more frequency bands with noise $\sim$ O(1) $\mu$K-arcmin upto 500 GHz will be extremely useful for such detections. \color{black} We need to also take into account the effect of CMB primordial fluctuations, internal motions and calibration etc for such detections. However, these details are beyond the scope of this paper. \color{black}

\section{Discussion}
\label{sec:discuss}
For the sake of simplicity, we have assumed a constant value of the ambient density in our radio galaxy evolutionary model. Realistically, the density increases with redshift, since it should scale with the critical density, thus implying smaller sized of radio \color{black}galaxy lobes \color{black} and higher radio power at larger redshifts, but also with rapid decline in radio power once the jet switches off. The net non-thermal SZ signal will be larger. Note, that for particular radio galaxies considered in this work (refer to Table \ref{tab:table2}), the estimates of radio power and $y_{\rm{NT}}$ will not differ significantly from those presented here.

The SZ effect from galaxy clusters can be used as cluster mass proxy in SZ surveys which aims at using cluster number counts and their spatial correlations to constrain cosmological parameters. However, this is crucially dependent on establishing an unbiased SZ distortion - cluster mass scaling.  The thermal SZ effect $y$-distortion is the dominant distortion for large clusters ($y\sim 10^{-5}-10^{-4}$ for clusters of virial mass $\sim$ $\times10^{14}M_\odot - 10^{15}M_\odot$). The non-thermal SZ signal from radio \color{black}lobes \color{black} inside the clusters can easily be a significant fraction of the thermal SZ from the cluster gas and cannot not be ignored. In the absence of observations at many different frequencies, one needs to model the non-thermal SZ (as in Table \ref{tab:table2}) and subtract it out from the total SZ distortion so as to have an unbiased estimate of the SZ from the cluster gas. However, if there are many observable frequencies, then the separation of the different components can be done by utilizing their unique spectral shapes as demonstrated in the previous section. Additionally, neglecting SZ distortions from radio \color{black}lobes \color{black} (inside clusters) would be bias the estimate of the Hubble Constant $H_0$ using a combination of SZ and XRay observations towards galaxy clusters. It is interesting to note that subtracting a contribution from radio \color{black}lobes \color{black} lowers the estimate of the net SZ distortion from the cluster gas, and pushes the the value of $H_0$ up, in the right direction.

Other than radio \color{black}galaxy lobes \color{black} residing within clusters, radio galaxies having \color{black}radio lobes \color{black} are ubiquitous in our Universe. In the previous sections, we have calculated the magnitude of distortion of the CMB spectrum from individual radio galaxy \color{black}lobes \color{black}. As a next step, we can consider an ensemble of radio \color{black}lobes \color{black} populating the universe and calculate the global averaged CMB distortions. A first calculation of the average $y$-distortion has been done in \cite{YSS1999} and \cite{M2001} assuming that the radio \color{black}galaxy lobes \color{black} reside inside the dark matter halos  with one-percent halo occupational efficiency.  The jet luminosity was assumed to be equal to the Eddington luminosity of central black hole of mass $M_{\rm BH}$ with $M_{\rm BH}=0.002M_{\rm halo}$. Moreover, \cite{M2001} calculated the angular power spectrum, $C_\ell$, of CMB distortions from unresolved radio galaxy \color{black}lobes \color{black}.  However, these initial efforts assumed the distortion to be non-relativistic $y$-distortion and concluded that a cosmological distribution of \color{black}radio lobes \color{black} with $t_{\rm J}$=$10^7$ yr to be severely constrained due to the COBE CMB spectral distortion limit \citep{COBE1996}. Our calculations suggest that in a $\Lambda$CDM universe with the current cosmological parameters, the global averaged $<y>\sim 10^{-6}$ for $t_{\rm J}=10^7$ yr.  Several improvements are in order to make progress - for example, the assumed jet luminosity of their model may be too high compared to the jet luminosity inferred from individual radio galaxies \citep{HC2020}. Also, the efficiency factor will not be a constant but a function of dark matter halo mass.  Our preliminary work on calculating the two point correlation functions of SZ fluctuations imply that the contribution from radio galaxies can be $\sim$ 10 percent of contribution from galaxy clusters. We will present the results for $<y>$ and $C_\ell$ for a cosmological distribution of radio \color{black}galaxy lobes \color{black} in a followup work.
\par

\section{Conclusions}
\label{sec:conclude}
We perform a detailed quantitative study of the non-thermal hot electrons in radio galaxy \color{black}lobes \color{black}, both during the lifetime of the radio jets and after the jets stop, as a potential source of non-thermal SZ distortion to CMB black body spectrum. Since the energetic particles inside the \color{black}radio lobes \color{black} cool via both synchrotron radiation and inverse Compton scattering, there is a correlation between emitted radio power and expected intensity of CMB distortion at any instant. Combining radio galaxy evolution models of \cite{KDA1997} with suitable modification for jet stopping as in \cite{N2010}, we are able to estimate the physical properties of the radio \color{black}lobes \color{black} at any instant of time.

We predict the value of $y_{\rm{NT}}$, given the observed size and radio power of a \color{black}galaxy lobe \color{black}, from our evolutionary model by taking into account the cooling of electrons of all energy. This is in contrast to previous works, for example \cite{CMBM2013}, in which the authors inferred the value of number density of electrons from radio and X-ray observations. \color{black} X-ray allows one to probe only a portion of the electron spectrum, whereas SZE,
being dependent on the total energy density of electrons, can provide a better
diagnostic of the electron spectrum. The lack of Faraday rotation from these galaxies can be due to deviation from equipartition, and may not directly tell us about $p_{\rm{min}}$. \color{black} Radio and X-ray observation constrain the spectrum of electrons at $\gamma>10^4$. The authors extrapolate the spectrum of electrons to lower energy electrons which are responsible for the SZ signal. We do no such extrapolation.
Further, we use the non-thermal spectrum of these relativistic particles to calculate the distortion on the CMB for a given size and radio power of the radio galaxy \color{black}lobe \color{black}. A summary of our predictions for 21 radio \color{black}galaxy lobes \color{black} is tabulated in Table \ref{tab:table2}. 
 The key points of this work are summarized below : 
\begin{itemize}
\item 
 Although the pressure inside a \color{black}radio lobe \color{black} changes as the  \color{black}lobe \color{black} expands, for a particular \color{black}lobe \color{black} size at any instant,  both the radio power and $y_{\rm{NT}}$ is independent of redshift. \color{black} This is a consequence of assuming the density of the surrounding medium to be independent of the redshift. As long as jet is turned on, radio power is just a function of the jet luminosity and not the jet lifetime i.e. the radio power from two sources with $t_{\rm J}=10^7$ and $10^8$ yr are the same when both  sources are relatively young. 
\item 
 The injected electrons, when the jet is on, cool efficiently via inverse Compton scattering especially at higher redshifts. After the jet shuts off, there is no more supply of energetic electrons. Therefore, as soon as the jet turns off, there is a steep fall of radio power as the energetic electrons which are responsible for radio emission have all cooled down. 
\item
For a fixed radio power, $y_{\rm{NT}}$ increases with decreasing \color{black}radio lobe \color{black} size as pressure is larger for smaller volume (which leads to the expansion of the \color{black}radio lobes \color{black}). The prospect of detecting non-thermal SZ from \color{black} radio lobes \color{black} increases if the \color{black} radio lobes \color{black} are young (or smaller in size) and/or jet the is active.
\item
A direct consequence of the above points is that radio galaxy \color{black}lobes \color{black} residing in cluster environments would be better potential targets for non-thermal SZ detection. Similarly, 
dead field radio galaxies, with large \color{black}radio lobe \color{black} sizes, are not favourable sources for non-thermal SZ detection.
\item
The analysis presented in this paper can successfully model the recent first detection of non-thermal SZ effect from the radio galaxy RG01\citep{MDCMSNW2017} . This gives us the confidence in predicting the non-thermal SZ distortion for a further sample of 20 radio galaxy sources (in Table \ref{tab:table2}) which can be targeted by upcoming ground based SZ searches.
\item
For a given intensity of distortion on the CMB, there is a degeneracy between $y_{\rm{NT}}$ and $g_{\rm{NT}}(x)$. In contrast to previous studies, we can predict $y_{\rm{NT}}$ from our galaxy evolution model which can then constrain the value of spectral function $g_{\rm{NT}}(x)$. 
\item
Radio and SZ detection of radio \color{black}lobes \color{black} can help determine the physical properties of energetic electrons inside the \color{black}radio galaxy lobes\color{black}.
For the non-thermal SZ detection from RG01, we are able to constrain the value of $p_{\rm{min}}$ (the lowest energy threshold of the electron spectrum) using SZ effect for the very first time. We find $p_{\rm{min}} \sim 1-2$ is needed to explain the observations. 
\item
We demonstrate that future CMB experiments, with higher frequency bands ($\gtrsim$ 300 GHz), are needed for differentiating non-thermal SZ from radio \color{black}lobes \color{black} from other SZ distortions (for example, kSZ distortions from clusters of galaxies).   In this respect CMB S4 \citep{CMBS42016}, PICO \citep{PICO2019}, CMB Bharat \citep{CMBBharat2018} would be most promising.\end{itemize}

\section*{Acknowledgements}
We acknowledge the use of computational facilities of Department of
 Theoretical Physics at Tata Institute of Fundamental Research,
 Mumbai. We acknowledge support of the Department of Atomic Energy, Government of India, under project no. 12-R\&D-TFR-5.02-0200. SKA is grateful for financial support from the Royal Society and Prof. Jens Chluba for the invitation to University of Manchester, during which a part of this work was done. BN wishes to thank Siddharth Malu for useful discussions. SM wishes to use this opportunity to recollect the very fond memory of his first meeting with late Sergio Colafrancesco when SM was an PhD student on his first trip outside his country and Sergio had driven his car to Roma Termini station to pick up SM so that young an Indian student does not get lost in the chaos of a new city. Finally, we would like to thank the referee for the detailed and thorough comments which have helped us in improving the manuscript.

\section{Data availability}
The data underlying in this article are available in this article.





\bibliographystyle{mnras}
\bibliography{cocoon}


\appendix




\bsp	
\label{lastpage}
\end{document}